%% file: 3c66b.tex
\documentclass[twocolumn]{aastex63}

\usepackage[utf8]{inputenc}
\usepackage{enumitem}

\usepackage[leftcaption]{sidecap}
\sidecaptionvpos{figure}{t}

\usepackage{graphicx}
\usepackage{hyperref}
\usepackage{color}

\newcommand{\src}{3C66B}

\newcommand{\Mc}{\mathcal{M}}
\newcommand{\eprise}{\texttt{enterprise}}
\newcommand{\eg}{e.\,g.}
\usepackage{amsmath}

\newcommand{\rev}[1]{{\color{black}{#1}}}

\usepackage{url}
\usepackage{natbib}

\defcitealias{Sudou2003}{S03}
\defcitealias{Iguchi2010}{I10}
\defcitealias{Jenet2004}{J04}
\defcitealias{11yrCW}{A19}

\received{May 14, 2020}
\revised{July 22, 2020}
\accepted{July 31, 2020}
\submitjournal{\apj}

\shorttitle{NANOGrav Multi-Messenger Searches: 3C66B}
\shortauthors{The NANOGrav Collaboration}

\hypersetup{colorlinks=true, citecolor=cyan, urlcolor = blue, linkcolor = blue}

\begin{document}

\title{Multi-Messenger Gravitational Wave Searches with Pulsar Timing Arrays: Application to 3C66B  Using the NANOGrav 11-year Data Set}

\input{3c66bauthors}

\correspondingauthor{Caitlin A. Witt$^{\color{magenta}\S}$}

\email{caitlin.witt@nanograv.org}

\begin{abstract}
When galaxies merge, the supermassive black holes in their centers may form binaries and, during the process of merger, emit low-frequency gravitational radiation in the process. In this paper we consider the galaxy \src, which was used as the target of the first multi-messenger search for gravitational waves. Due to the observed periodicities present in the photometric and astrometric data of the source, it has been theorized to contain a supermassive black hole binary. Its apparent 1.05-year orbital period would place the gravitational wave emission directly in the pulsar timing band. Since the first pulsar timing array study of \src, revised models of the source have been published, and timing array sensitivities and techniques have improved dramatically. With these advances, we further constrain the chirp mass of the potential supermassive black hole binary in \src\ to less than $(1.65\pm0.02) \times 10^9~{M_\odot}$ using data from the NANOGrav 11-year data set. This upper limit provides a factor of 1.6 improvement over previous limits, and a factor of 4.3 over the first search done. Nevertheless, the most recent orbital model for the source is still consistent with our limit from pulsar timing array data. In addition, we are able to quantify the improvement made by the inclusion of source properties gleaned from electromagnetic data \rev{over} `blind' pulsar timing array searches. With these methods, it is apparent that it is not necessary to obtain exact a priori knowledge of the period of a binary to gain meaningful astrophysical inferences.

\end{abstract}


\section{Introduction}\label{sec:intro}
\input{intro}

\section{Analysis Methods}\label{sec:analysis}
\input{methods}

\section{Results}\label{Results}
\input{Results}

\section{Discussion}\label{Discussion}
\input{Discussion}
\section{Conclusions}\label{Conclusion}
\input{Conclusions}

\acknowledgments

\textit{Author Contributions.}
We list specific contributions to this paper below.
CAW led the work on this paper, ran the GW searches, and led the development of the manuscript.
JS, SRT, SJV, and SBS provided guidance throughout the project and provided key development of the project motivation and scientific interpretation. 
JAE, SRT, PTB, SJV, and CAW designed and implemented the Bayesian search algorithms in \texttt{enterprise}.
JSH performed and interpreted the S/N simulations with \texttt{hasasia}. 
RDE performed initial literature reviews on \src. 
NJC, JSH, DLK, MTL, TJWL, MAM, CMFM, and DJN contributed valuable scientific comments.
NANOGrav data is the result of the work of dozens of people over the course of more than thirteen years.
ZA, KC, PBD, MED, TD, JAE, ECF, RDF, EF, PAG, MLJ, MTL, RSL, MAM, CN, DJN, TTP, SMR, PSR, RS, IHS, KS, JKS, and WZ developed the 11-year data set. 
All authors are key contributing members to the NANOGrav collaboration.

\textit{Acknowledgments.} SBS and CAW are supported for this work by NSF awards \#1458952 and \#1815664. The NANOGrav collaboration is supported by NSF Physics Frontier Center award \#1430284. CAW acknowledges support from West Virginia University through the Outstanding Merit Fellowship for Continuing Doctoral Students. 
SBS is a CIFAR Azrieli Global Scholar in the Gravity and the Extreme Universe program.
This research made use of the Super Computing System (Spruce Knob) at WVU, which is funded in part by the National Science Foundation EPSCoR Research Infrastructure Improvement Cooperative Agreement \#1003907, the state of West Virginia (WVEPSCoR via the Higher Education Policy Commission) and WVU. We acknowledge use of Thorny Flat at WVU, which is funded in part by the National Science Foundation Major Research Instrumentation Program (MRI) Award \#1726534 and WVU.
NANOGrav research at UBC is supported by an
NSERC Discovery Grant and Discovery Accelerator Supplement and by the Canadian Institute for Advanced Research.
MV and JS acknowledge support from the JPL RTD program. SRT was partially supported by an appointment to
the NASA Postdoctoral Program at JPL, administered by Oak
Ridge Associated Universities through a contract with NASA.
JAE was partially supported by NASA through Einstein Fellowship grants PF4-150120. Portions of this work performed
at NRL are supported by the Chief of Naval Research. The
Flatiron Institute is supported by the Simons Foundation. Portions of this research were carried out at the Jet Propulsion
Laboratory, California Institute of Technology, under a contract with the National Aeronautics and Space Administration. 
Data for this project were collected
using the facilities of the Green Bank Observatory and the
Arecibo Observatory. 
Green Bank Observatory is a facility of the National Science Foundation operated under cooperative agreement by Associated Universities, Inc. 
The Arecibo Observatory is a facility of the National Science Foundation operated under cooperative agreement by the University of Central
Florida in alliance with Yang Enterprises, Inc. and Universidad Metropolitana.
The National Radio Astronomy Observatory is a facility of the National Science Foundation operated under cooperative agreement by Associated Universities, Inc.
 This work made use of the online cosmology calculator tool \citep{cosmocalc}. We also acknowledge use of \texttt{numpy} \citep{numpy}, \texttt{scipy} \citep{scipy}, \texttt{matplotlib} \citep{matplotlib}, and \texttt{astropy} \citep{astropy}. This research has made use of the NASA/IPAC Extragalactic Database (NED), which is operated by the Jet Propulsion Laboratory, California Institute of Technology, under contract with the National Aeronautics and Space Administration. 

\bibliographystyle{aasjournal}
\bibliography{3c66b}

\end{document}

%% file: 3c66bauthors.tex
% DO NOT EDIT THIS FILE. EDITS WILL BE OVERWRITTEN.
% AUTO-GENERATED WITH make-aastex62-author-list.py
% FROM author_list_12yr_wb.txt, author_affil_and_orcid.txt, AND affil.txt
\author{Zaven Arzoumanian}
\affiliation{X-Ray Astrophysics Laboratory, NASA Goddard Space Flight Center, Code 662, Greenbelt, MD 20771, USA}
\author[0000-0003-2745-753X]{Paul T. Baker}
\affiliation{Department of Physics and Astronomy, Widener University, One University Place, Chester, PA 19013, USA}
\author{Adam Brazier}
\affiliation{Cornell Center for Astrophysics and Planetary Science and Department of Astronomy, Cornell University, Ithaca, NY 14853, USA}
\author[0000-0003-3053-6538]{Paul R. Brook}
\affiliation{Department of Physics and Astronomy, West Virginia University, P.O. Box 6315, Morgantown, WV 26506, USA}
\affiliation{Center for Gravitational Waves and Cosmology, West Virginia University, Chestnut Ridge Research Building, Morgantown, WV 26505, USA}
\author[0000-0003-4052-7838]{Sarah Burke-Spolaor}
\affiliation{Department of Physics and Astronomy, West Virginia University, P.O. Box 6315, Morgantown, WV 26506, USA}
\affiliation{Center for Gravitational Waves and Cosmology, West Virginia University, Chestnut Ridge Research Building, Morgantown, WV 26505, USA}
\author[0000-0003-0909-5563]{Bence Bécsy}
\affiliation{Department of Physics, Montana State University, Bozeman, MT 59717, USA}
\author[0000-0003-3579-2522]{Maria Charisi}
\affiliation{Theoretical AstroPhysics Including Relativity (TAPIR), MC 350-17, California Institute of Technology, Pasadena, California 91125, USA}
\author[0000-0002-2878-1502]{Shami Chatterjee}
\affiliation{Cornell Center for Astrophysics and Planetary Science and Department of Astronomy, Cornell University, Ithaca, NY 14853, USA}
\author[0000-0002-4049-1882]{James M. Cordes}
\affiliation{Cornell Center for Astrophysics and Planetary Science and Department of Astronomy, Cornell University, Ithaca, NY 14853, USA}
\author[0000-0002-7435-0869]{Neil J. Cornish}
\affiliation{Department of Physics, Montana State University, Bozeman, MT 59717, USA}
\author[0000-0002-2578-0360]{Fronefield Crawford}
\affiliation{Department of Physics and Astronomy, Franklin \& Marshall College, P.O. Box 3003, Lancaster, PA 17604, USA}
\author[0000-0002-6039-692X]{H. Thankful Cromartie}
\affiliation{University of Virginia, Department of Astronomy, P.O. Box 400325, Charlottesville, VA 22904, USA}
\author[0000-0002-1529-5169]{Kathryn Crowter}
\affiliation{Department of Physics and Astronomy, University of British Columbia, 6224 Agricultural Road, Vancouver, BC V6T 1Z1, Canada}
\author[0000-0002-2185-1790]{Megan E. DeCesar}
\affiliation{Department of Physics, Lafayette College, Easton, PA 18042, USA}
\author[0000-0002-6664-965X]{Paul B. Demorest}
\affiliation{National Radio Astronomy Observatory, 1003 Lopezville Rd., Socorro, NM 87801, USA}
\author[0000-0001-8885-6388]{Timothy Dolch}
\affiliation{Department of Physics, Hillsdale College, 33 E. College Street, Hillsdale, Michigan 49242, USA}
\author{Rodney D. Elliott}
\affiliation{Department of Astrophysical and Planetary Sciences, University of Colorado Boulder, Boulder, CO 80309, USA}
\author{Justin A. Ellis}
\affiliation{Infinia ML, 202 Rigsbee Avenue, Durham NC, 27701}
\author[0000-0002-2223-1235]{Robert D. Ferdman}
\affiliation{School of Chemistry, University of East Anglia, Norwich, NR4 7TJ, United Kingdom}
\author{Elizabeth C. Ferrara}
\affiliation{NASA Goddard Space Flight Center, Greenbelt, MD 20771, USA}
\author[0000-0001-8384-5049]{Emmanuel Fonseca}
\affiliation{Department of Physics, McGill University, 3600  University St., Montreal, QC H3A 2T8, Canada}
\author[0000-0001-6166-9646]{Nathan Garver-Daniels}
\affiliation{Department of Physics and Astronomy, West Virginia University, P.O. Box 6315, Morgantown, WV 26506, USA}
\affiliation{Center for Gravitational Waves and Cosmology, West Virginia University, Chestnut Ridge Research Building, Morgantown, WV 26505, USA}
\author[0000-0001-8158-638X]{Peter A. Gentile}
\affiliation{Department of Physics and Astronomy, West Virginia University, P.O. Box 6315, Morgantown, WV 26506, USA}
\affiliation{Center for Gravitational Waves and Cosmology, West Virginia University, Chestnut Ridge Research Building, Morgantown, WV 26505, USA}
\author[0000-0003-1884-348X]{Deborah C. Good}
\affiliation{Department of Physics and Astronomy, University of British Columbia, 6224 Agricultural Road, Vancouver, BC V6T 1Z1, Canada}
\author[0000-0003-2742-3321]{Jeffrey S. Hazboun}
\affiliation{University of Washington Bothell, 18115 Campus Way NE, Bothell, WA 98011, USA}
\author{Kristina Islo}
\affiliation{Center for Gravitation, Cosmology and Astrophysics, Department of Physics, University of Wisconsin-Milwaukee,\\ P.O. Box 413, Milwaukee, WI 53201, USA}
\author[0000-0003-1082-2342]{Ross J. Jennings}
\affiliation{Cornell Center for Astrophysics and Planetary Science and Department of Astronomy, Cornell University, Ithaca, NY 14853, USA}
\author[0000-0001-6607-3710]{Megan L. Jones}
\affiliation{Center for Gravitation, Cosmology and Astrophysics, Department of Physics, University of Wisconsin-Milwaukee,\\ P.O. Box 413, Milwaukee, WI 53201, USA}
\author[0000-0002-3654-980X]{Andrew R. Kaiser}
\affiliation{Department of Physics and Astronomy, West Virginia University, P.O. Box 6315, Morgantown, WV 26506, USA}
\affiliation{Center for Gravitational Waves and Cosmology, West Virginia University, Chestnut Ridge Research Building, Morgantown, WV 26505, USA}
\author[0000-0001-6295-2881]{David L. Kaplan}
\affiliation{Center for Gravitation, Cosmology and Astrophysics, Department of Physics, University of Wisconsin-Milwaukee,\\ P.O. Box 413, Milwaukee, WI 53201, USA}
\author[0000-0002-6625-6450]{Luke Zoltan Kelley}
\affiliation{Center for Interdisciplinary Exploration and Research in Astrophysics (CIERA), Northwestern University, Evanston, IL 60208}
\author[0000-0003-0123-7600]{Joey Shapiro Key}
\affiliation{University of Washington Bothell, 18115 Campus Way NE, Bothell, WA 98011, USA}
\author[0000-0003-0721-651X]{Michael T. Lam}
\affiliation{School of Physics and Astronomy, Rochester Institute of Technology, Rochester, NY 14623, USA}
\affiliation{Laboratory for Multiwavelength Astronomy, Rochester Institute of Technology, Rochester, NY 14623, USA}
\author{T. Joseph W. Lazio}
\affiliation{Jet Propulsion Laboratory, California Institute of Technology, 4800 Oak Grove Drive, Pasadena, CA 91109, USA}
\affiliation{Theoretical AstroPhysics Including Relativity (TAPIR), MC 350-17, California Institute of Technology, Pasadena, California 91125, USA}
\author[0000-0002-2034-2986]{Lina Levin}
\affiliation{Jodrell Bank Centre for Astrophysics, University of Manchester, Manchester, M13 9PL, United Kingdom}
\author{Jing Luo}
\affiliation{Department of Astronomy \& Astrophysics, University of Toronto, 50 Saint George Street, Toronto, ON M5S 3H4, Canada}
\author[0000-0001-5229-7430]{Ryan S. Lynch}
\affiliation{Green Bank Observatory, P.O. Box 2, Green Bank, WV 24944, USA}
\author[0000-0003-2285-0404]{Dustin R. Madison}
\affiliation{Department of Physics and Astronomy, West Virginia University, P.O. Box 6315, Morgantown, WV 26506, USA}
\affiliation{Center for Gravitational Waves and Cosmology, West Virginia University, Chestnut Ridge Research Building, Morgantown, WV 26505, USA}
\author[0000-0001-7697-7422]{Maura A. McLaughlin}
\affiliation{Department of Physics and Astronomy, West Virginia University, P.O. Box 6315, Morgantown, WV 26506, USA}
\affiliation{Center for Gravitational Waves and Cosmology, West Virginia University, Chestnut Ridge Research Building, Morgantown, WV 26505, USA}
\author[0000-0002-4307-1322]{Chiara M. F. Mingarelli}
\affiliation{Center for Computational Astrophysics, Flatiron Institute, 162 5th Avenue, New York, New York, 10010, USA}
\affiliation{Department of Physics, University of Connecticut, 196 Auditorium Road, U-3046, Storrs, CT 06269-3046, USA}
\author[0000-0002-3616-5160]{Cherry Ng}
\affiliation{Dunlap Institute for Astronomy and Astrophysics, University of Toronto, 50 St. George St., Toronto, ON M5S 3H4, Canada}
\author[0000-0002-6709-2566]{David J. Nice}
\affiliation{Department of Physics, Lafayette College, Easton, PA 18042, USA}
\author[0000-0001-5465-2889]{Timothy T. Pennucci}
\altaffiliation{NANOGrav Physics Frontiers Center Postdoctoral Fellow}
\affiliation{National Radio Astronomy Observatory, 520 Edgemont Road, Charlottesville, VA 22903, USA}
\affiliation{Institute of Physics, E\"{o}tv\"{o}s Lor\'{a}nd University, P\'{a}zm\'{a}ny P. s. 1/A, 1117 Budapest, Hungary}
\author[0000-0002-8826-1285]{Nihan S. Pol}
\affiliation{Department of Physics and Astronomy, West Virginia University, P.O. Box 6315, Morgantown, WV 26506, USA}
\affiliation{Center for Gravitational Waves and Cosmology, West Virginia University, Chestnut Ridge Research Building, Morgantown, WV 26505, USA}
\author[0000-0001-5799-9714]{Scott M. Ransom}
\affiliation{National Radio Astronomy Observatory, 520 Edgemont Road, Charlottesville, VA 22903, USA}
\author[0000-0002-5297-5278]{Paul S. Ray}
\affiliation{Space Science Division, Naval Research Laboratory, Washington, DC 20375-5352, USA}
\author[0000-0002-7283-1124]{Brent J. Shapiro-Albert}
\affiliation{Department of Physics and Astronomy, West Virginia University, P.O. Box 6315, Morgantown, WV 26506, USA}
\affiliation{Center for Gravitational Waves and Cosmology, West Virginia University, Chestnut Ridge Research Building, Morgantown, WV 26505, USA}
\author[0000-0002-7778-2990]{Xavier Siemens}
\affiliation{Department of Physics, Oregon State University, Corvallis, OR 97331, USA}
\affiliation{Center for Gravitation, Cosmology and Astrophysics, Department of Physics, University of Wisconsin-Milwaukee,\\ P.O. Box 413, Milwaukee, WI 53201, USA}
\author[0000-0003-1407-6607]{Joseph Simon}
\affiliation{Jet Propulsion Laboratory, California Institute of Technology, 4800 Oak Grove Drive, Pasadena, CA 91109, USA}
\affiliation{Theoretical AstroPhysics Including Relativity (TAPIR), MC 350-17, California Institute of Technology, Pasadena, California 91125, USA}
\author[0000-0002-6730-3298]{Ren\'{e}e Spiewak}
\affiliation{Centre for Astrophysics and Supercomputing, Swinburne University of Technology, P.O. Box 218, Hawthorn, Victoria 3122, Australia}
\author[0000-0001-9784-8670]{Ingrid H. Stairs}
\affiliation{Department of Physics and Astronomy, University of British Columbia, 6224 Agricultural Road, Vancouver, BC V6T 1Z1, Canada}
\author[0000-0002-1797-3277]{Daniel R. Stinebring}
\affiliation{Department of Physics and Astronomy, Oberlin College, Oberlin, OH 44074, USA}
\author[0000-0002-7261-594X]{Kevin Stovall}
\affiliation{National Radio Astronomy Observatory, 1003 Lopezville Rd., Socorro, NM 87801, USA}
\author[0000-0002-1075-3837]{Joseph K. Swiggum}
\altaffiliation{NANOGrav Physics Frontiers Center Postdoctoral Fellow}
\affiliation{Department of Physics, Lafayette College, Easton, PA 18042, USA}
\author[0000-0003-0264-1453]{Stephen R. Taylor}
\affiliation{Department of Physics and Astronomy, Vanderbilt University, 2301 Vanderbilt Place, Nashville, TN 37235, USA}
\author[0000-0002-4162-0033]{Michele Vallisneri}
\affiliation{Jet Propulsion Laboratory, California Institute of Technology, 4800 Oak Grove Drive, Pasadena, CA 91109, USA}
\affiliation{Theoretical AstroPhysics Including Relativity (TAPIR), MC 350-17, California Institute of Technology, Pasadena, California 91125, USA}
\author[0000-0003-4700-9072]{Sarah J. Vigeland}
\affiliation{Center for Gravitation, Cosmology and Astrophysics, Department of Physics, University of Wisconsin-Milwaukee,\\ P.O. Box 413, Milwaukee, WI 53201, USA}
\author[0000-0002-6020-9274]{Caitlin A. Witt$^{\color{magenta}\S}$}
\affiliation{Department of Physics and Astronomy, West Virginia University, P.O. Box 6315, Morgantown, WV 26506, USA}
\affiliation{Center for Gravitational Waves and Cosmology, West Virginia University, Chestnut Ridge Research Building, Morgantown, WV 26505, USA}

\author[0000-0001-5105-4058]{Weiwei Zhu}
\affiliation{National Astronomical Observatories, Chinese Academy of Science, 20A Datun Road, Chaoyang District, Beijing 100012, China}

\collaboration{1000}{The NANOGrav Collaboration}
%\altaffiliation{Author order alphabetical by surname}
\noaffiliation

%% file: intro.tex
Continuous gravitational waves (GWs), defined by single-source cyclic GW emission, are expected to arise from the supermassive black hole binaries (SMBHBs) that form during a galaxy merger. When a SMBHB evolves such that it emits GWs in the microhertz to nanohertz GW band (orbital periods of weeks to several decades), a sufficiently massive and/or nearby SMBHB may be detectable by pulsar timing arrays (PTAs; e.g., \citealt{11yrCW}) (hereafter \citetalias{11yrCW}). 

While GWs from individual sources in the PTA regime have been sought after in multiple works \citep{5yrcw, cpta_1909_cw, Jenet2004} through a variety of methods, none have been detected. However, numerous advances have been made in the field of pulsar timing. As PTA experiments gain longer time baselines and higher cadences and the numbers of millisecond pulsars grows, sensitivity to GW sources increases.
The notable ongoing PTA programs in the world include the European PTA, Parkes PTA, and the North American Nanohertz Observatory for Gravitational Waves (NANOGrav) \citep[\eg][respectively]{EPTA,PPTA,11yr_data}. Altogether, these PTAs time approximately 100 pulsars to high precision with the goal of GW detection, among other endeavors \citep[\eg][]{IPTA-planet-paper,IPTA-clock-paper}. 

In addition, sophisticated GW detection methods have been developed to detect quadrupolar continuous-wave signals in the data of coordinated timing arrays \citep[\eg][]{EPTA-CW-paper,PPTA-cw-paper,11yrCW}. However, past analyses that used the most up-to-date methods have used `blind' detection methods; that is, the software did not consider any binary model information gained from electromagnetic data to directly benefit the search. In comparison, most works that do target \rev{limits on} specific sources using electromagnetic information have used smaller data sets consisting of a single pulsar \rev{with} as a periodogram approach \citep{Jenet2004, cpta_1909_cw},  \rev{or have been derived from the stochastic GW background \citep{zhu_2019}} rather than the full GW analysis pipeline. Here, we have combined these methods in the first search of this type, where we used the entire NANOGrav array of pulsars and full GW search analysis, while incorporating electromagnetic data to conduct a more informed search for GWs from our test source, \src.

Since the report of a hypothesized orbital motion in the core of the galaxy \src\ by \citet{Sudou2003} (hereafter \citetalias{Sudou2003}), it has been an ideal test case for searches for GWs from SMBHBs. Using long-baseline interferometry, the authors found apparent elliptical motions in 3C66B's radio core, modeling this motion as the gyration of the jet nozzle due to an orbit-induced precession of the \rev{smaller} black hole's jet. \citetalias{Sudou2003} proposed a period and chirp mass for the binary of ${1.05 \pm 0.03}$ years and ${1.3\times10^{10}~ M_\odot}$, respectively. Given the relatively small redshift of the galaxy ($z=0.02126$), a binary with those properties would be emitting gravitational radiation well within the sensitivity of pulsar timing arrays (PTAs).

As such, \src\ has long been a prime candidate for continuous GW detection. It was the first object targeted for continuous wave detection, as reported by \citet{Jenet2004} (hereafter \citetalias{Jenet2004}), in which seven years of Arecibo timing data from PSR B1855+09 \citep{Kaspi} was used to search the Fourier domain timing residuals (commonly referred to as a Lomb-Scargle periodogram), using harmonic summing \citep{press+1992}, for a GW signal consistent with the binary period modeled by \citetalias{Sudou2003}. With these methods, they did not see evidence of a significant signal, and were able to place an upper limit of ${7\times10^9~M_\odot}$ on the chirp mass of the system at a binary eccentricity of $e=0$.

Since the study of \citetalias{Jenet2004},  \citet{Iguchi2010} (hereafter \citetalias{Iguchi2010}), reported a 93-day variability in the active galactic nucleus's \rev{millimeter} light that was interpreted as likely due to doppler boosting of a relativistic outflow that is modulated by orbital motion (its period differs due to geometric effects). The new model assumed the 1.05-year orbital period from \citetalias{Sudou2003}, but predicted an updated chirp mass of ${7.9\times10^8~M_\odot}$, almost a full order of magnitude lower than the upper limit set by \citetalias{Jenet2004}. 
%\fixme{other binarity evidence?}

\rev{\src\ was also one of the objects targeted by \citet{zhu_2019}, which used a novel approach to test 3C66B indirectly by using the source to predict the GW background strength implied by this source's existence. They concluded that the \citetalias{Iguchi2010} model produced GW backgrounds that were larger than are currently probed by pulsar timing arrays, implying that the source was not likely to be a binary with parameters as proposed by \citetalias{Iguchi2010}.}

The work reported here presents a Bayesian cross-validation framework in which we use 3C66B's binary parameter measurements as priors for our continuous wave search. Our search has resulted in the most stringent \rev{direct} GW derived limit to date on the chirp mass of 3C66B's candidate SMBHB. We also test, more generically within our search framework, what sensitivity improvements can be gained by knowing the GW frequency of a target to increasingly good precision. 

Therefore, we have quantified the improvement made by searching for GWs from a specific source, including cases where the orbital period is only known with large error or not known at all.

Note that because \citetalias{Jenet2004} used only one pulsar in their study, they would have been unable to perform a formal experiment to detect \src, as the use of one pulsar precludes the ability to demonstrate the quadrupolar signature that is unique to the influence of gravitational waves. Thus, our study here is the first formal targeted detection experiment for \src\ using a pulsar timing array.

This paper is laid out as follows: in Section~\ref{sec:analysis}, we describe our data, mathematical model, and software pipeline.
In Section~\ref{Results} we report the detection Bayes factor and chirp mass upper limit for \src, as well as results for new test methods. In Section~\ref{Discussion} and Section~\ref{Conclusion}, we present our conclusions as well as discuss implications for future detection prospects of this and other SMBHBs.

%% file: methods.tex
\subsection{Pulsar Timing and Electromagnetic Data}\label{sec:data}

We make use of the NANOGrav 11-year Data Set \citep{11yr_data}, which provides high precision timing of 45 millisecond pulsars. Only the 34 pulsars with baselines of at least 3 years are used for GW detection analyses \citep{11yr_gwb}. We describe slight differences in the use of the data set in this work as compared to other papers in Section~\ref{sec:analysis}. However, the majority of the data are treated similarly to \citetalias{11yrCW}. Due to the 11-year timing baseline, the data set is most sensitive to binaries with orbital periods of less than a decade.

The electromagnetic data we incorporate into our models are mainly derived from \citetalias{Sudou2003} and \citetalias{Iguchi2010}, as well as the location from the NASA/IPAC Extragalactic Database (NED)\footnote{\label{nedfoot}The NASA/IPAC Extragalactic Database (NED) is operated by the Jet Propulsion Laboratory, California Institute of Technology, under contract with the National Aeronautics and Space Administration, and can be accessed at \url{https://ned.ipac.caltech.edu/}}. 
These values are summarized in Table \ref{tab:model_vals}. The right ascension, declination, and luminosity distance are taken as constants throughout the analysis, as the PTA sensitivity to sky location and distance is much lower than any associated errors.
For consistency with earlier work, we take the luminosity distance of \src\ to be 85 Mpc, as in \citetalias{Sudou2003}. Therefore, all calculations use $H_0 = 75~\mathrm{km~s^{-1}~Mpc^{-1}}$. Note that minor differences in the distance due to different reports of redshift or $H_0$ cause only a small fractional variation in the results. If the fractional change in the luminosity distance is defined as 

\begin{equation}
    d_{85} \equiv \left(\frac{d_L}{85~ \mathrm{Mpc}}\right),
\end{equation}
any GW strain limit can be converted to the reader's preferred distance by multiplying the strain by $d_{85}$, and $\Mc$ limits by multiplying by $d_{85}^{3/5}$.
{
\renewcommand{\arraystretch}{1.5}
\begin{table}
    \scriptsize
    \centering
    \caption{GW Model Values and Uncertainties}
    \begin{tabular}{ccc}
    \hline
         Parameter  & Value & Reference\\
         \hline
         \hline
         Chirp Mass ($\Mc$) & $7.9 ^{+3.8}_{-4.5} \times 10^8 \mathrm{M}_\odot$& \citetalias{Iguchi2010}\\

         GW frequency ($f_{\rm{GW}})$ & $60.4 \pm 1.73$ nHz &\citetalias{Sudou2003}\\
         Redshift ($z$) & 0.02126 & \rev{\citet{huchra}}\\
%         $d_L$ & 85.8 Mpc & NED$^{\footnotemark[0][\ref{nedfoot}]}$\\
         RA & 02h 23m 11.4112s &\rev{ \citet{fey}}\\
         Dec & \rev{+42d 59m 31.384s} & \rev{\citet{fey}}\\
         GW strain ($h$) & $7.3^{+6.8}_{-5.8} \times 10^{-15}$& \citetalias{Sudou2003, Iguchi2010}\\
         \hline
    \end{tabular}
    \label{tab:model_vals}
\end{table}
}

\subsection{Signal Model}

We use the methods presented in \citetalias{11yrCW} for the generation of expected pulsar timing residuals influenced by a signal from a continuous GW from a circular SMBHB. While we will not present the full derivation, we will summarize below the relevant equations needed to follow our analysis on the NANOGrav data and refer the reader to \citetalias{11yrCW} for more detail. Note that throughout this section, equations are written in natural units (where $G=c=1$).

Pulsar timing residuals describe the deviation of an observed pulse arrival time from that predicted from a model based on spin, astrometric, interstellar delay, and if needed, binary parameters of the pulsar. These are the basic data product that we use to search for GWs, which will not be included in the pulsar's timing model. 
\rev{A vector of timing residuals ($\delta t$) that is fit without a GW} for each pulsar is modeled as 

\begin{equation}
    \delta t = M\epsilon + n_{\rm white} + n_{\rm red} + s~,
\end{equation}
where $M$ is the design matrix, which describes the timing model, and $\epsilon$ is a vector of the linearized timing model parameter offsets from the best fit solution. In other words, the timing model, which was originally derived without the presence of a GW, must now be adjusted.
We write a vector describing the white noise in the data as $n_{\rm white}$, and the same for the red noise, $n_{\rm red}$, which is correlated over long timescales. The noise terms are described in more detail in
Section \ref{analyses}.

The signal $s$ can be derived as follows.
For a GW source whose sky location is described by polar and azimuthal angles $\theta$ and $\phi$, the strain induced by the emitted GWs is written in terms of two polarizations as

\begin{equation}
h_{a b}(t, \hat{\Omega})=e_{a b}^{+}(\hat{\Omega}) h_{+}(t, \hat{\Omega})+e_{a b}^{ \times}(\hat{\Omega}) h_{ \times}(t, \hat{\Omega})~,
\end{equation}
where $h_{+,\times}$ are the polarization amplitudes and $e_{a b}^{+, \times}$ are the polarization tensors, which we write in the solar system barycenter (SSB) frame as 

\begin{align}
    e_{a b}^{+}(\hat{\Omega}) &= \hat{m}_{a} \hat{m}_{b}-\hat{n}_{a} \hat{n}_{b}\\
    e_{a b}^{ \times}(\hat{\Omega}) &= \hat{m}_{a} \hat{n}_{b}+\hat{n}_{a} \hat{m}_{b}, 
\end{align}
\citep{wahlquist}. In these equations, we define $\hat{\Omega}$ as a unit vector pointing from the GW source to the SSB, written as

\begin{equation}
    \hat{\Omega}=-\sin \theta \cos \phi \hat{x}-\sin \theta \sin \phi \hat{y}-\cos \theta \hat{z}~.
\end{equation}
We define the vectors $\hat{m}$ and $\hat{n}$ as
\begin{align}
    \hat{m} & =\sin \phi \hat{x}-\cos \phi \hat{y},\\
    \hat{n} &= -\cos \theta \cos \phi \hat{x}-\cos \theta \sin \phi \hat{y}+\sin \theta \hat{z}~.
\end{align}

The pulsar's response to the GW source is described by the antenna pattern functions (\citealt{sesana_vecchio,ellis2012, taylor2016} and references therein)

\begin{align}
    F^{+}(\hat{\Omega}) &=\frac{1}{2} \frac{(\hat{m} \cdot \hat{p})^{2}-(\hat{n} \cdot \hat{p})^{2}}{1+\hat{\Omega} \cdot \hat{p}}, \\ 
    F^{ \times}(\hat{\Omega}) &=\frac{(\hat{m} \cdot \hat{p})(\hat{n} \cdot \hat{p})}{1+\hat{\Omega} \cdot \hat{p}}~, 
\end{align}
where $\hat{p}$ is a unit vector pointing from the Earth to the pulsar. 
%\rev{We can also define the quantity cos($\mu$) as $-\hat{\Omega} \cdot \hat{p}$}

Finally, we write the signal $s$ induced by the GW, as seen in pulsar's residuals, as 
\begin{equation}
    s(t, \hat{\Omega})=F^{+}(\hat{\Omega}) \Delta s_{+}(t)+F^{ \times}(\hat{\Omega}) \Delta s_{ \times}(t).
\end{equation}
Here, $\Delta s_{+, \times}$ represents the difference between the signal induced at the Earth (the Earth term) and that at the pulsar (the pulsar term), and can be written as
\begin{equation}
    \Delta s_{+, \times}(t)=s_{+, \times}\left(t_{p}\right)-s_{+, \times}(t)
\end{equation}
where $t$ is the time at which the GW passes the SSB and $t_p$ is the time the GW passes the pulsar.\footnote{This definition is occasionally written as the negative of the right side of the equation here, e.g., $s_{+, \times}(t)-s_{+, \times}\left(t_{p}\right)$ as in \citetalias{11yrCW}. This is resolved with a change of convention in the definition of the GW antenna pattern, as we have done here; thus all results are consistent between these works.} These times can be related from geometry by 
\begin{equation}
    t_{p}=t-L(1+\hat{\Omega} \cdot \hat{p})
\end{equation}
where $L$ is the distance to the pulsar.

For a circular binary at zeroth post-Newtonian order, $s_{+, \times}$ is given by \citep{wahlquist, lee, corbin}

\begin{align} 
\begin{split}
    s_{+}(t)=& \frac{\mathcal{M}^{5 / 3}}{d_{L} \omega(t)^{1 / 3}}\left[-\sin 2 \Phi(t)\left(1+\cos ^{2} i\right) \cos 2 \psi\right.\\ &-2 \cos 2 \Phi(t) \cos i \sin 2 \psi], \\ s_{ \times}(t)=& \frac{\mathcal{M}^{5 / 3}}{d_{L} \omega(t)^{1 / 3}}\left[-\sin 2 \Phi(t)\left(1+\cos ^{2} i\right) \sin 2 \psi\right.\\ &+2 \cos 2 \Phi(t) \cos i \cos 2 \psi],
\end{split}
\end{align}
where $i$ is the inclination angle of the SMBHB, $\psi$ is the GW polarization angle, $d_L$ is the luminosity distance to the source, and $\Mc$ is the chirp mass, which is related to the two black hole masses as
\begin{equation}
    \mathcal{M} = \frac{\left(m_{1} m_{2}\right)^{3 / 5}}{\left(m_{1}+m_{2}\right)^{1 / 5}}.
\end{equation}
It is important to note that $\Mc$ and $\omega,$ in this case, refer to the observed redshifted values. 

For a circular binary, we relate the orbital angular frequency to the GW frequency with $\omega_{0}=\pi f_{\mathrm{GW}}$, where $\omega_{0}=\omega\left(t_{0}\right)$. For this work, as in \citetalias{11yrCW} we define $t_0$ as the last MJD in the 11-year data set (MJD 57387). The orbital phase and frequency of the SMBHB are given by
\begin{align}
    \Phi(t)&=\Phi_{0}+\frac{1}{32} \mathcal{M}^{-5 / 3}\left[\omega_{0}^{-5 / 3}-\omega(t)^{-5 / 3}\right], \\ \omega(t)&=\omega_{0}\left(1-\frac{256}{5} \mathcal{M}^{5 / 3} \omega_{0}^{8 / 3} t\right)^{-3 / 8},
\end{align}
where $\Phi_0$ and $\omega_0$ are the initial orbital phase and frequency. As in \citetalias{11yrCW}, we use the full expression for $\omega(t)$ to maintain consistency across runs, as this form is needed to model the signal at the higher frequencies sampled in some runs, as described in Section \ref{sss:f_test}.

\subsection{Software and Analyses}\label{analyses}

In this work, we make use of NANOGrav's GW detection package, \eprise\footnote{\url{https://github.com/nanograv/enterprise}}, an open-source code written fully in Python that 
contains a built-in interface with the pulsar timing data and noise models required to perform Bayesian GW analysis \citep[limits and detection]{11yr_gwb}.
Basic algorithms for Bayesian continuous wave analysis are described in detail in a number of past works \citep[\eg][]{ellis+13,ellis+cornish}

Using \eprise, we can use a priori constraints on a binary system, which come from electromagnetic observation (for instance, the period of 3C66B) to set priors on GW parameters that are derived from the binary model.  
%These priors define the assumed confidence interval (e.g., measured values and errors) for the parameters that are used to construct a continuous GW signal. 
Within \eprise\ we can easily add these priors to the timing model and noise model to obtain a full model of the signal. We then perform Markov-Chain Monte Carlo (MCMC) methods implemented in  \texttt{PTMCMCSampler}\footnote{\url{https://github.com/jellis18/PTMCMCSampler}} to find the posterior distribution for each of the free parameters. For `blind' continuous wave (CW) searches as in \citetalias{11yrCW}, we typically set uninformative priors, which are uniform across the allowed range of values, for the binary system's parameters, such as sky location, frequency, mass, and distance to the source. Thus, the methods here could be considered a ``targeted'' search by our use of informed priors. 

For instance, in the simplest treatment of \src, a specific binary model has been hypothesized, with measurements and associated unknowns in the mass, mass ratio, and orbital frequency \citepalias[\eg][]{Sudou2003,Iguchi2010}. We can use these electromagnetically constrained parameters, in addition to knowledge of the location of this object on the sky, to restrict our priors.\footnote{Note that our restricted priors might not always be Gaussian; in some cases, electromagnetic observations of a source may produce a model that contains greater complexity than Gaussian error bars. In such cases, non-Gaussian priors must be used. The functionality exists in \eprise\ for studies that would require such a setup. As an example, if cyclic flux variability is observed, the period of variability might represent the fundamental orbital frequency, a harmonic, or even a resonance, requiring a multi-valued prior. In our analysis, the reported errors on binary masses from \citetalias{Iguchi2010} were asymmetric, and thus for some analyses, our chirp mass prior required an asymmetric distribution.}

Assuming a SMBHB with a circular orbit, a continuous GW signal can be characterized by eight of the following nine parameters:

\begin{equation}
    \{\theta,\phi,f_{\rm{GW}},\Phi_0,\psi,i,\Mc,d_{\rm L},{h_0}\},
\end{equation}
which represent the GW source's:
\begin{itemize}[noitemsep, nolistsep]
    \item position on the sky $(\theta, \phi)$;
    \item GW frequency, related to the orbital frequency at some reference time $(f_{\rm GW})$;
    \item orbital phase at some reference time $(\Phi_0)$;
    \item GW polarization angle $(\psi)$;
    \item orbital inclination $(i)$;
    \item chirp mass ($\Mc$);
    \item luminosity distance ($d_{\rm L}$);
    \item strain amplitude $(h_0)$, which is related to the chirp mass, GW frequency, and luminosity distance .
\end{itemize}

The ninth parameter is redundant, as the strain amplitude $h_0$ can be defined by

\begin{equation}
    h(t)=F_{+} h_{+}+F_{\times} h_{\times}=A h_{0} \cos \left(\Phi(t)-\Phi_{0}\right)
\end{equation}
\citep{Sathyaprakash}, where 

\begin{align*}
\begin{split}
    A=&\left(A_{+}^{2}+A_{\times}^{2}\right)^{1 / 2}\\
    A_{+}=&~\frac{1}{2} F_{+}\left(1+\cos i^{2}\right)\\ A_{\times}=&~F_{\times} \cos i,
\end{split}
\end{align*}

and can be related to other physical parameters by

\begin{equation}
    h_0 = \frac{2 {\Mc^{5/3}}(\pi f_{\mathrm{GW}})^{2/3}}{{d_L}}. 
\end{equation}
Since the strain is entirely determined by $\Mc$, $f_{\rm GW}$, and $d_L$, a limit on $h_0$ based on a PTA search can be translated into constraints on these source parameters.
Since the uncertainties on $\theta$, $\phi$, and $d_L$ are much smaller than the PTA sky localization accuracy, by targeting a specific source with a known position and redshift, we can set these parameters as constant values, and therefore reduce the number of search parameters to five.

In all runs, there is also a set of free parameters associated with each pulsar included in the PTA which are varied in the analysis. First of these is the pulsar distance, which has a Gaussian prior in all cases. In pulsars where the distance is reported in \citet{dist}, the Gaussian is defined using the recognized distance and the associated error. For the remaining pulsars, the Gaussian prior is \rev{set to a fiducial} 1.0$\pm$0.2 kpc, \rev{which is consistent with the distribution of distances and uncertainties obtained from \citet{dist}. Although this range does not necessarily encompass the actual distances to most of these pulsars, it works as a proxy value, and the choice of this value does not affect our results.}
%\rev{This assumption is, for most pulsars, consistent to three sigma with the dispersion measure distance derived from both \citet{ymw} and \citet{ne2001}.} 
As in \citetalias{11yrCW}, this assumption can be seen to hold in the posteriors for these pulsars, as the prior is returned in all cases, meaning \rev{this analysis} cannot inform on the distances for these pulsars. This is expected, as these pulsars are largely those with shorter observation baselines, which are influencing the PTA to a smaller degree.
%\rev{, and would be unlikely to show a strong signal in the pulsar term.}
\rev{The recovered pulsar distances also affect the GW frequency difference between the Earth and the pulsar, which therefore will be related to the chirp mass. When a wide range of chirp mass values are allowed by the data, the uncertainty in the pulsar distances is not significant to the final result of the search. Additionally, for small chirp masses, for even a large change in the distance to the pulsar, the change induced in the GW frequency at the pulsar is well below the resolution limit of the PTA ($1/T_{obs}$). This angular frequency at the pulsar can be calculated as 
\begin{equation}
    \omega_{p,0} = \omega_0 \left(1 + \frac{256}{5} \Mc^{5/3} \omega_0^{8/3}d_p(1+\hat{\Omega} \cdot \hat{p})\right)^{-3/8},
\end{equation}
where $d_p$ is the distance from the Earth to the pulsar.}

Also included is the GW phase at the pulsar. While this quantity could be calculated geometrically from the other parameters, including it as a search parameter mitigates potential issues sampling the complex parameter space, which arise due to the large uncertainty on the distances to the pulsars compared to the GW wavelength. 

As is standard for these types of analyses, (e.g., \citealt{11yr_gwb,11yrCW}) the white noise of each pulsar (described as EFAC, EQUAD, and ECORR) is held fixed. The power spectral density of the pulsar intrinsic red noise is modeled as 
\begin{equation}
    P = \frac{A_{\rm red}^2}{12 \pi^2} \left( \frac{f}{f_{\rm yr}}\right)^{-\gamma} \mathrm{yr}^3, 
\end{equation}
where $A_{\rm red}$ (the red noise amplitude) and $\gamma$ (the red noise spectral index) are also allowed to vary in each pulsar in our Markov-Chain Monte Carlo simulation. Here, $f_{\rm yr}$ is $1/(1 \rm {yr})$ in Hz. To assist the sampler, empirical distributions of the red noise parameters were made from single pulsar noise run posteriors and used to create jump proposals. These determine how steps in the MCMC are taken through generating proposed samples, and were added to significantly improve sampling and decrease burn-in time for our analyses. For a more detailed description, see Appendix A of \citetalias{11yrCW}. 

Our treatment of the red noise in one pulsar, J0613$-$0200, required additional noise modeling. As described in \citetalias{11yrCW}, this pulsar possesses extra unmodeled noise processes that, in the 11-year continuous wave search, presented as an increase in strain upper limit at a frequency of 15 nHz. In this work, this manifested as poor sampling in the CW parameters, particularly in $f_{\rm GW}$. Because of this poor sampling, the $f_{\rm GW}$ parameter would periodically get stuck near this frequency. \rev{Due to this pulsar's location relative to \src, which places it among the ten pulsars  with the highest antenna pattern response amplitudes, it is important to find a robust solution to these issues rather than remove the pulsar from the analysis.} To mitigate this effect, we applied more sophisticated noise modeling techniques to allow the red noise to deviate from the typical power-law, with corresponding jump proposals to assist sampling. The noise model that was chosen is a $t$-process spectrum, which allows for `fuzziness' in the typical power-law spectrum by scaling the power spectral density by a variable factor for each frequency. This model is 
created by generalizing the typical Gaussian process prior to a Student's $t$-distribution. This process will be discussed in more depth in \citet{adv_noise_mod}\rev{, and, due to increasingly complex data, will likely become more typical in future analyses.} 

\begin{figure} %%skymap 
    \centering
    \includegraphics[width=\columnwidth]{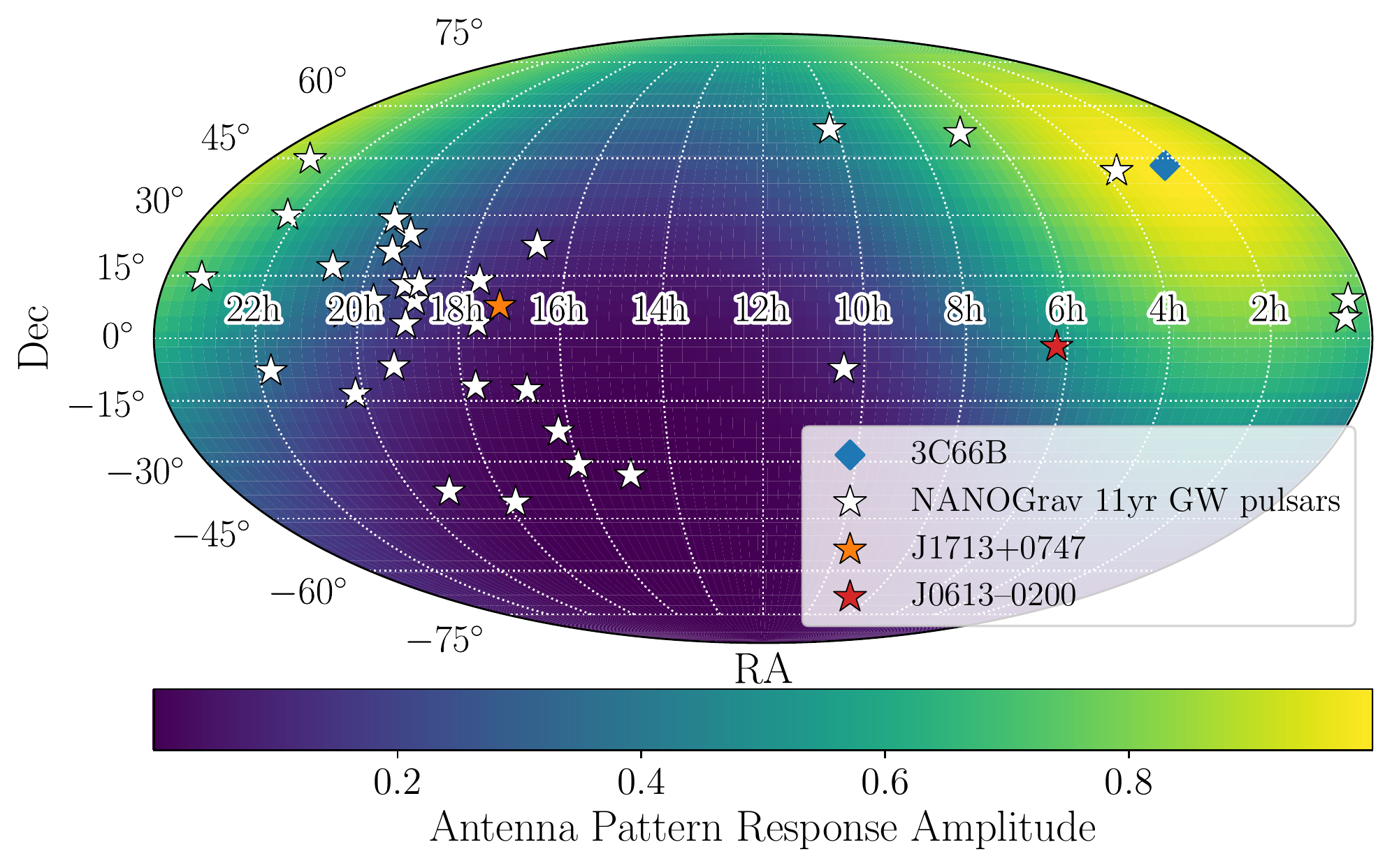}
    \caption{Sky map depicting the antenna pattern response amplitude ($F_\times^2+F_+^2$) due to a GW located at the sky position of \src. Also plotted are the locations of the 34 pulsars used in GW analyses of the NANOGrav 11-year data set, with the two pulsars in need of special attention noted with separate colors.}
    \label{fig:skymap}
\end{figure}

Even with this model, poor sampling in the $f_{\rm GW}$ parameter still occurred, and can be attributed to unmodeled noise due to changes in the dispersion measure of pulsar J1713+0747, caused by variations in the interstellar medium along the line of sight \citep{dip2, slices}. While this pulsar is NANOGrav's most sensitive in general, it is not particularly sensitive to \src, as shown in Figure \ref{fig:skymap}, and thus excluding it did not significantly effect the upper limit on target \src. As such, this pulsar was removed from our search.

The above procedure is used for all \eprise\ runs as described in detail in the next subsection.

\subsection{Four Distinct Tests}
We constructed several separate set-ups for \eprise\ for the purpose of testing distinct hypotheses. The purpose of each of these, and the difference in procedures within \eprise, is described below.

\subsubsection{Detection}\label{detection}
To determine if a CW from \src\ is detected, we conduct an \eprise\ search using a single frequency, with a value corresponding to the 1.05-year orbital period for a circular binary, making the final set of search parameters 

\begin{equation}\label{eq:set}
    \{\Phi_0,\psi,i,\Mc\}.
\end{equation}
\rev{Due to the frequency resolution of the PTA, which is defined by the timing baseline, it is reasonable to set a parameter with errors of this magnitude (Table \ref{tab:model_vals}) to a constant value. However, we will explore the relaxation of this assumption in later sections.}
Note that the \citetalias{Iguchi2010} and \citetalias{Sudou2003} models make assumptions about the electromagnetic data which may or may not be correct; our model simply tests the presence of a SMBHB in this system at a period of 1.05\,years.

The detection prior on $\Mc$ is log-uniform in the range ${10^7~\mathrm{to}~10^{10}~M_\odot}$, and is sampled in log-space. This prior is convenient for calculating Bayes factors as a measure of detection significance, using the Savage-Dickey formula \citep{dickey},

\begin{equation}
    	\mathcal{B}_{10} \equiv \frac{\text{evidence}[\mathcal{H}_1]}{\text{evidence}[\mathcal{H}_0]} = \frac{p(h_0 = 0|\mathcal{H}_1)}{p(h_0 = 0|\mathcal{D},\mathcal{H}_1)},
\end{equation}
Here, $\mathcal{H}_1$ is the model with a GW signal plus individual pulsar red noise, and $\mathcal{H}_0$ is the model with individual pulsar red noise only. The prior and posterior volumes at $h_{0}=0$ are $p\left(h_{0}=0 | \mathcal{H}_{1}\right)$ and $p\left(h_{0}=0 | \mathcal{D}, \mathcal{H}_{1}\right)$, respectively. We are able to apply the Savage-Dickey formula because these models are nested ($\mathcal{H}_0$ is $\mathcal{H}_{1}$ where $h_{0}=0$), and $p\left(h_{0}=0 | \mathcal{D}, \mathcal{H}_{1}\right)$ is approximated as the fraction of quasi-independent samples in the lowest-amplitude bin of a histogram of $h_0$. The error in the Bayes factor is computed as
\begin{equation}
    \sigma=\frac{\mathcal{B}_{10}}{\sqrt{n}},
\end{equation}
where $n$ is the number of samples in the lowest amplitude bin. This process is done once the samples in GW strain are calculated from the directly sampled parameters. In the detection analyses, the red noise amplitude is sampled with a matching prior (log-uniform in $A_{\rm red}$). All other GW parameters are searched with a uniform prior.

\subsubsection{Upper Limits}\label{upper}

To set an upper limit on the chirp mass of \src, we again conduct an \eprise\ search using a single frequency, with a value corresponding to the 1.05-year orbital period, making the final parameter set as in the previous section (Equation \ref{eq:set}).
However, in contrast with the case for detection, the upper limit prior on $\Mc$ is uniform (rather than log-uniform) meaning the prior set on the $\rm{log}_{10}\Mc$ exponentially increases over the range $\{7,10\}$. This is done as an astrophysically reasonable prior, as we expect SMBHBs to lie anywhere in this mass range, while still allowing for efficient sampling. Additionally, this prior choice allows the derived upper limit to be as conservative as possible by allowing a higher proportion of high chirp mass samples, and be independent from the choice of lower prior bound. In the upper limit analyses, the red noise amplitude is sampled with a matching prior (uniform in $A_{\rm red}$). Upper limits are taken to be the value of the 95th percentile of the posterior distribution. Following the approach of \citet{11yr_gwb}, we calculate the error on upper limit calculations as 

\begin{equation}
    \sigma=\frac{\sqrt{x(1-x) / N_{s}}}{p\left(h_{0}=h_{0}^{95 \%} | \mathcal{D}\right)},
\end{equation}
where $x=0.95$ and $N_s$ is the number of effective samples in the chain, which is estimated by dividing the total number of samples by the autocorrelation length of the chain. 

\subsubsection{Frequency Prior Testing}\label{sss:f_test}
In addition to the tests described above of the \citetalias{Sudou2003} and \citetalias{Iguchi2010} models, where the GW frequency is fixed to discrete values as in other continuous wave searches (\citetalias{11yrCW}; \citealt{5yrcw}), it is also crucial to test frequencies within the confidence region of these values. For this aim, we have developed methods to directly sample in $f_{\rm{GW}}$. These include specialized parameter groupings and jump proposals to help the sampler move through the more complex parameter space. Using these techniques, we are able to obtain an upper limit from the $\Mc$ posteriors for a variety of frequency priors from various \eprise\ setups. 

When searching over GW frequency, a log-uniform chirp mass prior is used, and the samples are re-weighted \rev{during upper limit calculations} to modify the prior choice from a uniform-in-log distributions of masses to a uniform-in-linear distributions of masses, the latter of which is more common in upper-limit analyses by virtue of insensitivity to the lower sampling boundary. This both assists with sampling and maintains a consistent prior on the GW strain, which is not directly sampled. To match the $\Mc$ prior, a log-uniform prior is used on $A_{\rm red}$. Since we are no longer fixing $f_{GW}$ to a single value, our final parameter set for these searches was 
\begin{equation}
    \{\Phi_0,\psi,i,\Mc,f_{\rm{GW}}\},
\end{equation}

In addition, we also chose to limit our GW frequency prior to a range of 1--100 nHz, rather than the 1--300 nHz used in \citetalias{11yrCW}. Besides the PTA's insensitivity at these high frequencies, we expect a source to remain in these frequency bins for very little time, with residence timescales as small as months, so their detection prospects are minimal \citep{review,hasasia}.

Using the three priors shown in Table \ref{tab:mc_freq}, we are able to find re-weighted upper limits for a variety of scenarios. 
These include:
\begin{enumerate}[noitemsep, nolistsep]
    \item The GW frequency is known, and set to a single value
    \item The GW frequency is known with large errors, and the error region is searched over
    \item The GW frequency is not known or has  \rev{significant uncertainty}, and the entire PTA sensitivity band is searched over.
\end{enumerate}
Then, we examined the change in re-weighted chirp mass upper limit as a function of frequency prior width. In addition to allowing for possible errors in the orbital period measured by \citetalias{Sudou2003}, these widened priors allow us to test the feasibility of this process on a less constrained source. Additionally, if there was any significant frequency evolution in the source, a signal would still have the chance to be detected in either of these setups. In addition to a single value and a uniform prior across the PTA sensitivity bandwidth, we also use 10 times the uncertainty on the predicted frequency \rev{as an example of a search with significant uncertainty}. We also bin the samples of the widest $f_{\rm{GW}}$ search to interpolate between these \rev{three individual prior widths}. The results of this examination are described in Section \ref{Results}, and are summarized in Figure \ref{fig:Mc_plot}

{
\renewcommand{\arraystretch}{1.5}
\begin{table}%%mc_freq
    \centering
    \caption{\rev{Frequency Prior Testing Weighted Upper Limits}}
    \begin{tabular}{crc}
    \hline
         Scenario & $f_{\rm{GW}}$ Prior  & Weighted $\mathcal{M}$ Upper Limit ($10^9 M_\odot$)\\
         \hline
         \hline
         1 & Constant & 1.57$\pm$ 0.02 \\

         2 & 10$\sigma$  &  1.54 $\pm$  0.01\\

         3 & Log-Uniform & 8.68 $\pm$0.07\\
         \hline
    \end{tabular}

    \label{tab:mc_freq}
\end{table}
}

\subsubsection{Test of a Specific Binary Model}\label{iguchi_methods}

To directly test the consistency of the model presented in \citetalias{Iguchi2010} with the NANOGrav data, we create priors for an \eprise\ run corresponding to the values presented (see the first line of Table \ref{tab:iguchi}). For $f_{\rm{GW}}$, we are able to use a Gaussian prior, where the error on the measured value from \citetalias{Iguchi2010} directly corresponds to the standard deviation of the prior. However, $\Mc$ has uneven error bars, so a more complicated prior is needed. Here, we fit a skewed normal distribution to the reported value and error, and construct a skewed normal prior based on this distribution, and also keep a log-uniform prior on $A_{\rm red}$. Therefore, the final parameter set for this search was 

\begin{equation}
    \{\Phi_0,\psi,i,\Mc,f_{\rm{GW}}\}.
\end{equation}

To analyze the amount of information gained between the prior and posterior models, we employed the Kullback–Leibler divergence \citep{KLtest}. We calculate this information gain in bits between the posterior $p(x|d)$ and the prior $p(x)$ as 
\begin{equation}
    D_{\mathrm{KL}}(P \| Q)=\int_{-\infty}^{\infty} p(x|d) \log \left(\frac{p(x|d)}{p(x)}\right) d x.
\end{equation}
This is done for the distributions for both $\Mc$ and $f_{\rm{GW}}$. To maintain consistency between forms of the posterior and the prior, we fit a skewed normal distribution to both posteriors to directly compare to the prior. 

%% file: Results.tex
\rev{The results discussed in this section can be reproduced, and the MCMC data examined, using code provided for the reader's convenience.\footnote{\url{https://github.com/nanograv/11yr_3c66b}}}

\subsection{Detection} \label{Detection}

\begin{figure}
    \centering
    \includegraphics[width=\columnwidth]{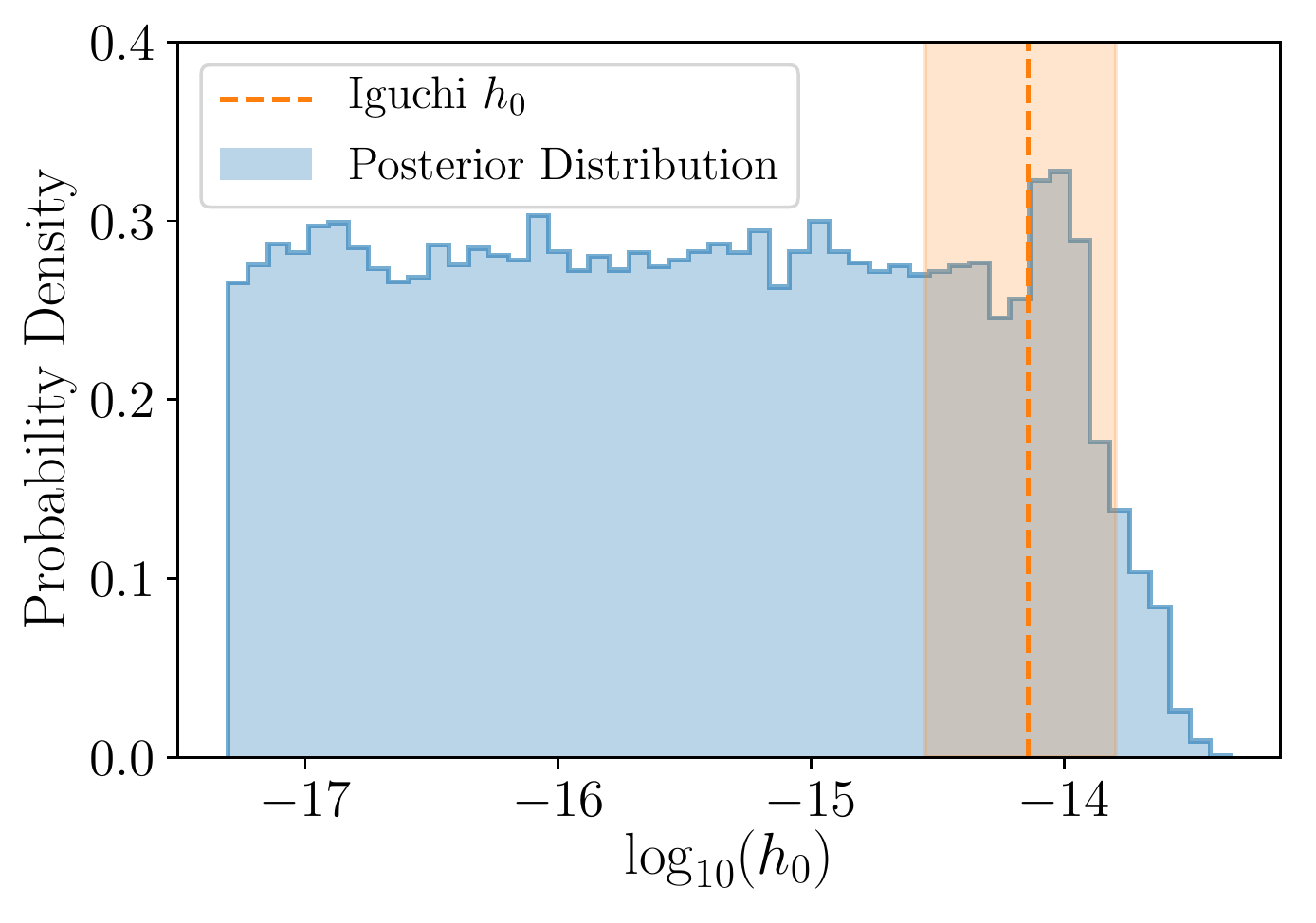}
    \caption{Posterior for the detection analysis described in Section \ref{Detection} (blue). The vertical orange region describes the area of parameter space where a signal with the parameters found by \citetalias{Iguchi2010} would lie. While the upper end of the parameter space is ruled out, there is clearly no value that is preferred by the sampler.}
    \label{fig:detection_plot}
\end{figure}

Using the setup for a detection run as described in Section \ref{detection}, we find no evidence for a GW signal from \src. We calculate a Savage-Dickey Bayes factor of $\mathcal{B}_{10} = 0.74 \pm 0.02$. Therefore, there is no evidence for the detection of a GW signal in the data. The posterior for this run is plotted in Figure \ref{fig:detection_plot}.

\subsection{Upper Limits}

\begin{figure} %%upper
    \centering
    \includegraphics[width=\columnwidth]{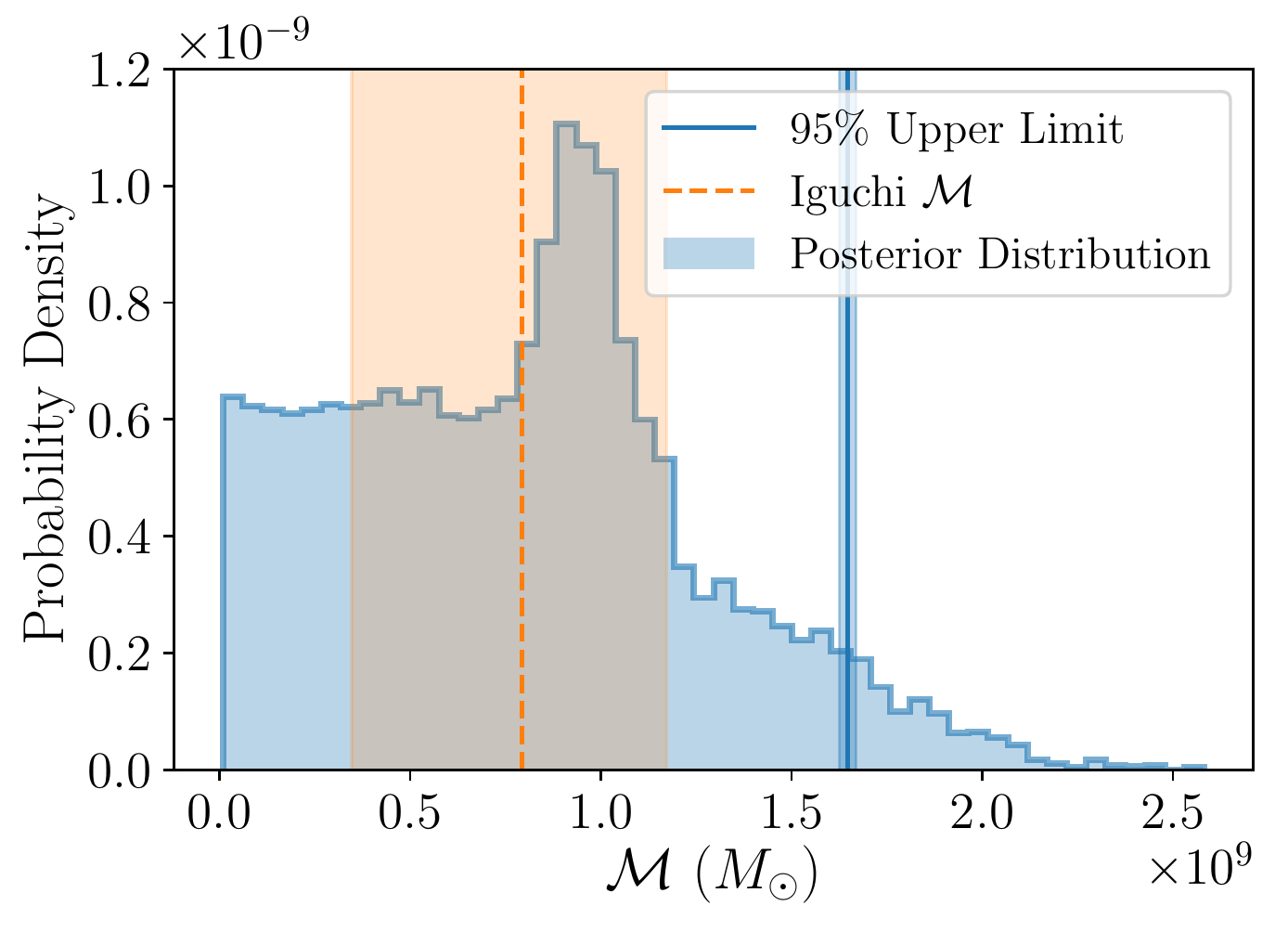}
    \caption{The chirp mass posterior histogram is plotted in blue, with a vertical line depicting the 95\% upper limit. Shown in orange is the chirp mass upper limit of \citetalias{Iguchi2010}, with the shaded region representing the error on the value. With these methods, the \citetalias{Iguchi2010} mass estimate is impossible to rule out. We also note that the peak in the posterior at $1\times10^{\rev{9}} M_\odot$ is not statistically significant.}
    \label{fig:Upper_plot}
\end{figure}

As no GW signal is detected from \src, we set upper limits on the chirp mass using the procedure described in Section \ref{upper}. Using the constant-value frequency prior at 60.4 nHz (corresponding to the 1.05-year orbital period), we set a 95\% upper limit of $(1.65\pm0.02) \times 10^9~{M_\odot}$ for $\Mc$ of the SMBHB in \src. This value corresponds to a strain of $(2.47 \pm 0.05) \times 10^{-14}$. To compare, the expected strain of the model in \citetalias{Iguchi2010} is $(7.2^{+6.8}_{-5.8}) \times 10^{-15}$. As can be seen in Figure \ref{fig:Upper_plot}, while we achieve a factor of 4.3 improvement over the limit set by \citetalias{Jenet2004}, we cannot rule out the \citetalias{Iguchi2010} model. The posterior distribution of samples does include a peak at about $1 \times 10^9 ~{M_\odot}$, which is within the error region for the chirp mass calculated from \citetalias{Iguchi2010}. However, this peak is not statistically significant, and is able to be traced to a single pulsar, J1909$-$3744. \rev{By examining the posterior distributions constructed from samples corresponding to this peak, we find structure in the GW phase posterior at J1909$-$3744 that does not occur for any other pulsar.} This likely occurs due to covariances between the model and sinusoidal behavior caused by noise processes in the data \rev{as a real GW signal would be recovered by more than one pulsar}. Therefore, this peak in the posterior is not indicative of a signal, and our upper limit can be considered robust. We will note that the upper limit listed can be calculated for the reader's preferred distance using the transformation described in Section \ref{sec:data}.

\subsection{Frequency Prior Testing}\label{ftest}

As described in Section \ref{sss:f_test}, we also performed tests to quantify how much our upper limits might improve if we have constrained (through electromagnetic observation) the orbital frequency of the target. While for \src\ the orbital frequency is assumed to be known to within small errors, for other targets, a frequency may not be known or be only poorly constrained. This test provides a sense of how well the period must be constrained to provide effective sensitivity gains for a GW search.

\begin{figure}
    \centering
    \includegraphics[width=\columnwidth]{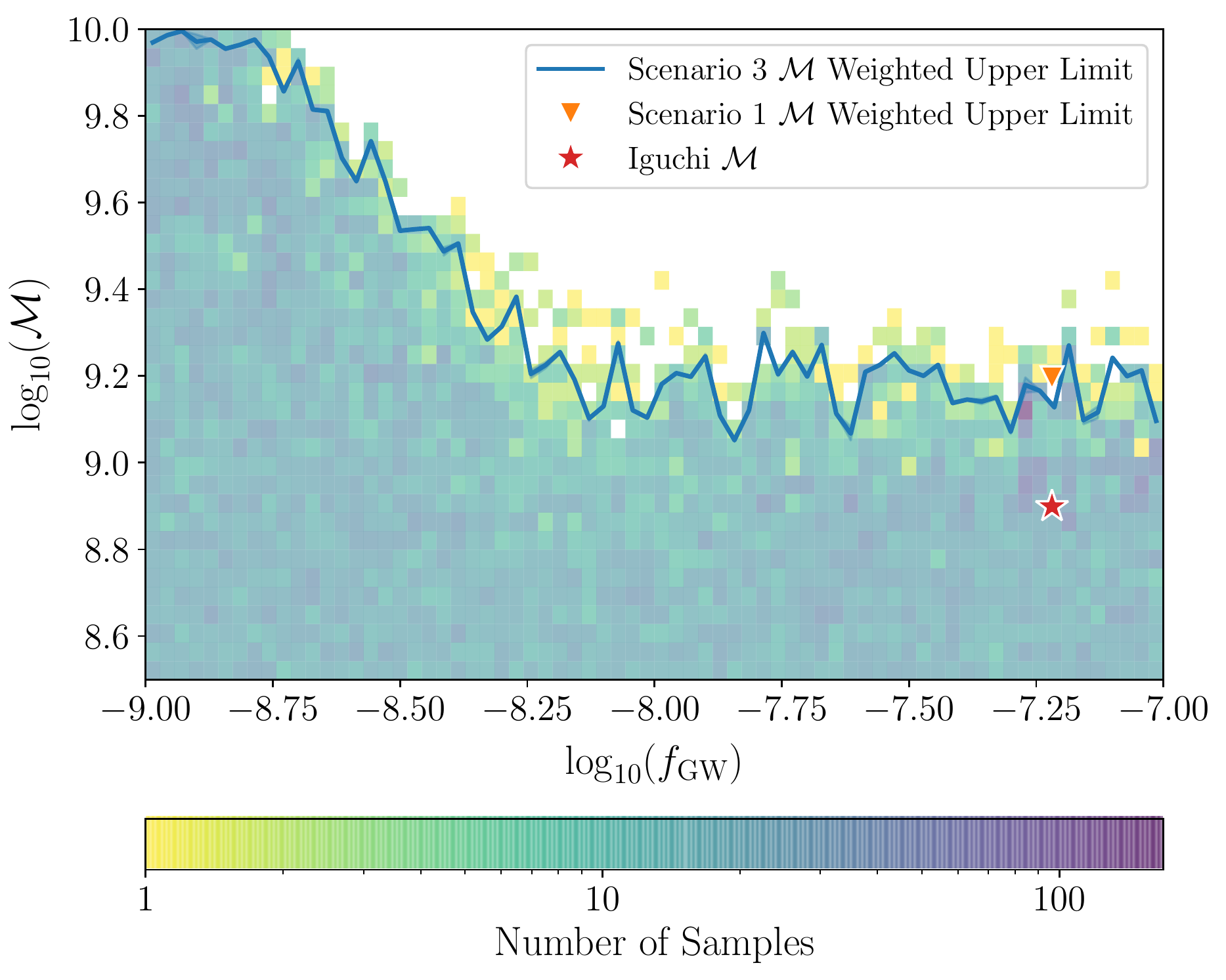}
    \caption{2D histogram of samples in the log-uniform prior setup. Also plotted is the weighted 95\% upper limit for each frequency bin (blue) from the scenario 3 setup. The white area indicates the section of parameter space ruled out by our search. It is clear from the uniform distribution of samples across all frequency and mass channels that all sampling issues have been resolved. This uniform distribution also makes clear that there is no indication of a signal at the distance and sky location of \src. We only plot the upper half of the parameter space in $\Mc$ to resolve more detail. Below $\mathrm{log}_{10}\Mc$ = 8.5, all sampling is uniformly distributed, identically to the upper half of the figure. For comparison, the scenario 1 weighted upper limit (orange triange) and \citetalias{Iguchi2010} chirp mass estimate (red star) are also shown. }
    \label{fig:2d_hist}
\end{figure}

Using the three scenarios described above, we are able to characterize the change in re-weighted upper limit between the setups. The result of the log-uniform prior search over the entire frequency band is summarized with Figure \ref{fig:2d_hist}. The white area represents the area of $\mathcal{M}$-$f_{\rm GW}$ parameter space ruled out in this analysis. From the uniformity of the samples over the parameter space, it is clear there are no sampling issues. This is due to the improved sampling methods described in Section \ref{analyses}. The weighted 95$\%$ upper limit is plotted for each frequency bin, allowing us to quantify for which frequencies we are the most sensitive to \src. \rev{We note that for the very lowest frequencies, the upper limit is dependent upon the choice of prior, as the search cannot rule out any of the prior range. }

\begin{figure} %%mc_plot 
    \centering
    \includegraphics[width=\columnwidth]{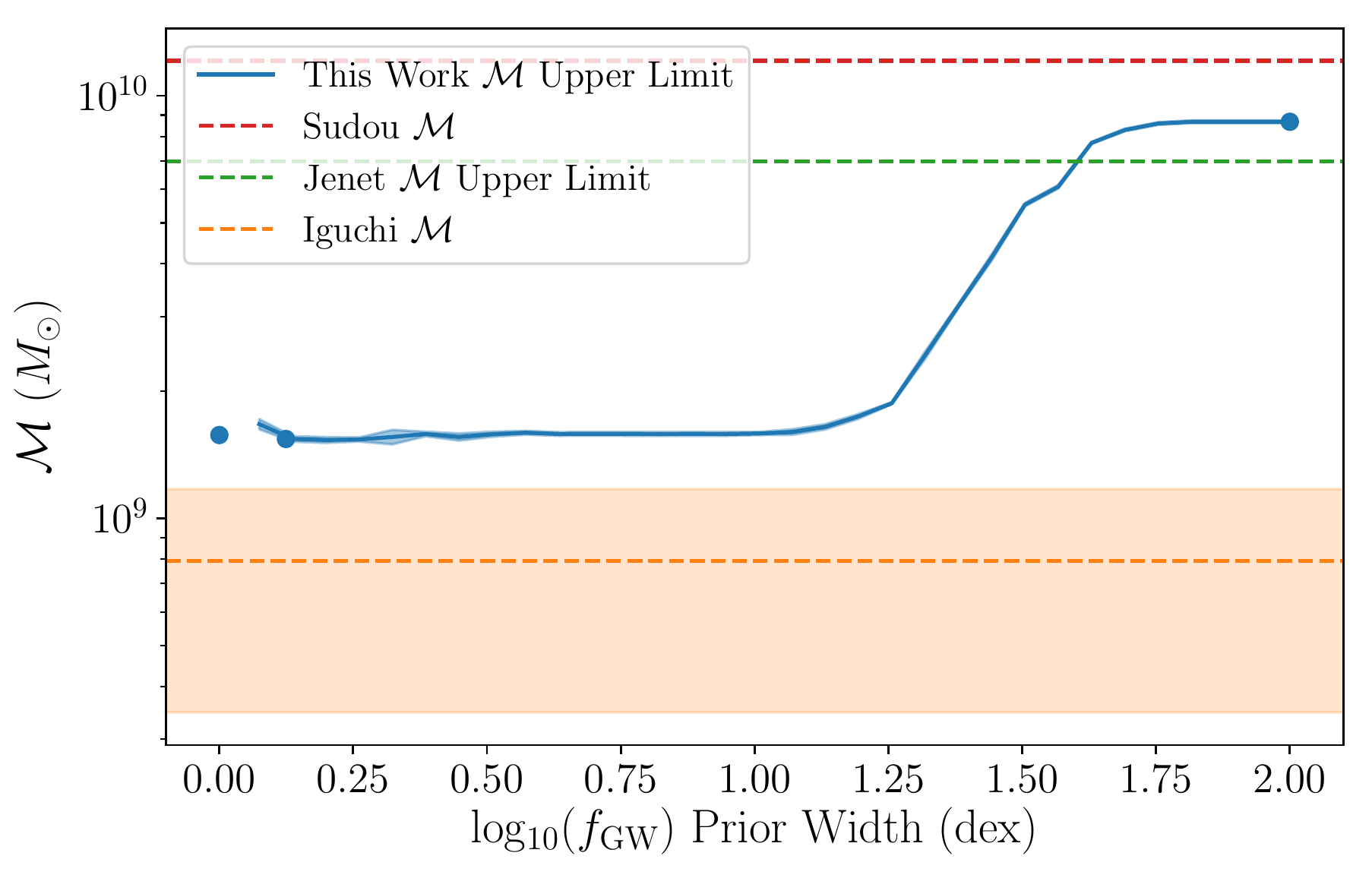}
    \caption{Chirp mass upper limits plotted with respect to frequency prior width (blue). Also shown as horizontal lines are previous upper limits set by \citetalias{Sudou2003} \rev{(red)}, \citetalias{Jenet2004} \rev{(green)}, and \citetalias{Iguchi2010} (orange), from top to bottom. Shaded regions describe error bars on the quoted limit. It is clear that none of these upper limits rule out that of \citetalias{Iguchi2010}. However, this figure accentuates the fact that when a period is known to less than 1 order of magnitude of precision, the limits on the target's mass improve by nearly one order of magnitude; that is, while the tightest prior produces the lowest upper limit, moderately wide priors also produce similar results, indicating that perfect orbital models would not be necessary to perform such a search on other systems. It is not until the prior spans approximately an order of magnitude that sensitivity is lost. Also plotted for comparison are the weighted upper limits for each of the three separate runs.}
    \label{fig:Mc_plot}
\end{figure}

In addition to the three runs described above, it was also possible to infer the upper limit that would be derived from a run with a frequency prior width between those of the three separate runs. To do this, we bin the samples in the scenario 3 (widest $f_{\rm GW}$ prior) run to keep only a certain range of frequencies and recalculate the weighted upper limit for this subset. \rev{These bins increase symmetrically in log space about the value of $f_{\rm{GW}}$ reported by \citetalias{Sudou2003}, from a log space full-width of 0 dex (a constant) until the upper bound reaches $f_{\rm{GW}}= 100$ nHz. After this, only the lower bound expands to reach a full log space full-width of 2 dex (essentially, 2 orders of magnitude in linear space).} The weighted upper limits calculated from these binned samples are plotted in Figure \ref{fig:Mc_plot}. 

Also plotted in Figure \ref{fig:Mc_plot} are the upper limits from the three individual runs. From the consistency of these points with the calculated curve, it is clear that this technique is robust. Additionally, this shows the feasibility of searching over $f_\mathrm{GW}$, as the results are consistent with those calculated for both an individual frequency and a small range.

As can be seen in Figure \ref{fig:Mc_plot} and Table \ref{tab:mc_freq}, there is nearly an order of magnitude difference in the upper limits derived from frequency varied runs of different prior widths. \rev{Because the upper limits at the very lowest frequencies are dependent upon the prior choice, the difference seen here is a lower limit.} However, from the curve calculated from binned samples, we see that this increase does not begin until about one order of magnitude in frequency space about the \citetalias{Iguchi2010} value is included.
This implies that a targeted search such as this is worthwhile even without exact orbital information, as long as the frequency is known to within an order of magnitude.  

\subsection{Test of a Specific Binary Model}\label{iguchi_results}

\begin{figure*}
    \centering
    \includegraphics[width = \textwidth]{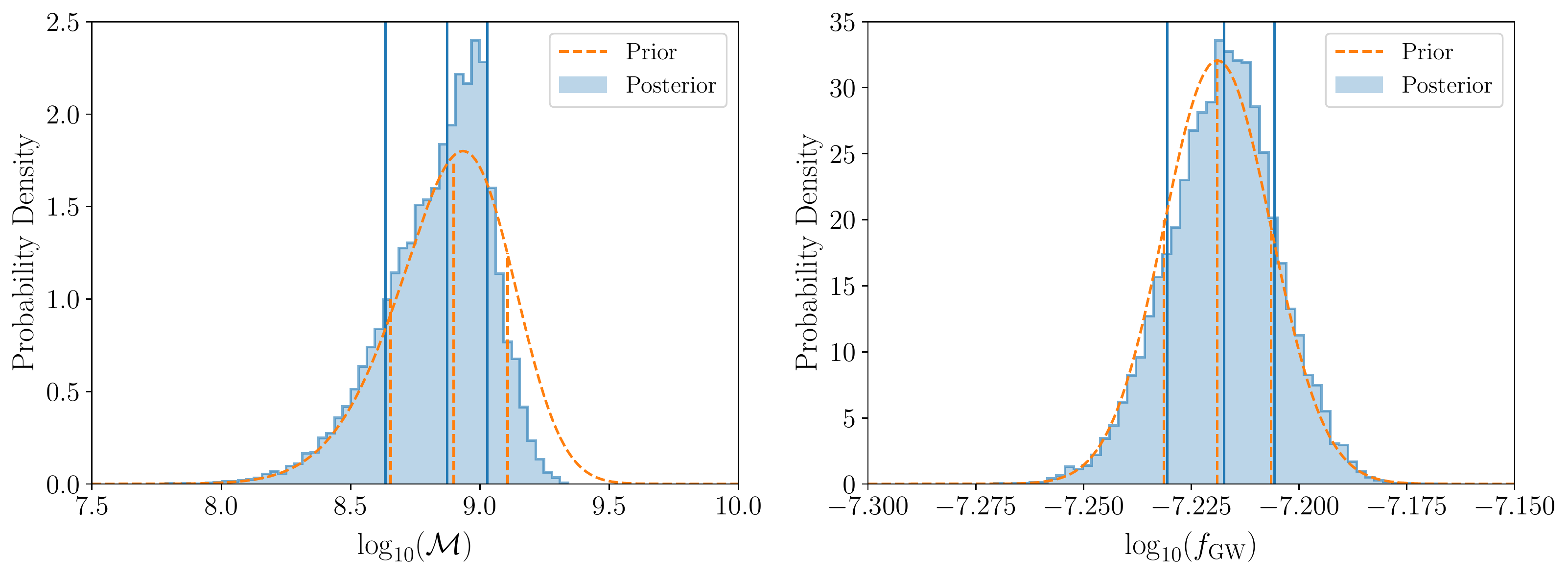}
    \caption{Posteriors (blue) and priors (orange) for the direct test of the model presented in \citetalias{Iguchi2010}. Vertical bars mark the \rev{16.86, 50, and 84.13 percentiles of each}, to represent the 1$\sigma$ error bars.}
    \label{fig:iguchi_test}
\end{figure*}

To directly test our sensitivity to a GW from the model of \src\ proposed in \citetalias{Iguchi2010}, we directly test priors as described in Section \ref{iguchi_methods}. In Figure \ref{fig:iguchi_test}, we can \rev{compare} the prior and posterior for both $f_{\rm{GW}}$ and $\Mc$. These \rev{distributions} are quantified in Table \ref{tab:iguchi}, where the error on the posterior values are calculated with the percentiles of the posterior distribution corresponding to 1$\sigma$ error bars. The values of $f_{\rm{GW}}$ are consistent with those of the prior, but for $\Mc$, we are able to significantly lower the upper bound on the value, effectively ruling out part of the high mass region of the model.

Additionally, we report the information gained between the posterior and the prior as described in Section \ref{iguchi_methods}. The differences in the distributions for $f_{\rm{GW}}$ produce a KL divergence of 0.0096, while those of the $\Mc$ distributions produce a KL divergence of 0.2597. While neither of these values is large, it is clear that much more information is gained about the chirp mass of \src\ from this model test than the GW frequency.

{
\renewcommand{\arraystretch}{1.5}
\begin{table}%%iguchi test
    \centering
    \caption{\rev{Model Testing Prior and Posterior Values}}
    \begin{tabular}{ccc}
    \hline
         &log(Frequency) & log(Chirp Mass)\\
         \hline
         \hline

         Iguchi (Prior) & $-7.219 \pm0.012$& $8.90^{+0.21}_{-0.24}$\\

         This Work (Posterior)  &  \rev{$-7.217^{+0.012}_{-0.013}$}& $8.87^{
         +0.16}_{-0.24}$\\

         \hline
    \end{tabular}

    \label{tab:iguchi}
\end{table}
}

%% file: Discussion.tex
To provide context for the upper limit on \src\ set in this work, we can compare to the limits set in \citetalias{11yrCW}, which do not have the benefit of electromagnetic constraints (i.\,e.\ a `blind' search). This comparison will allow us to estimate the improvement in sensitivity gained by including electromagnetic data over a typical blind search. By comparing our \rev{strain upper limit of $(2.47 \pm 0.05) \times 10^{-14}$} to the sensitivity curve in Figure 3 of \citetalias{11yrCW}, \rev{where the strain upper limit at the nearest searched frequency is $5.3\times 10^{-14}$ nHz,} we \rev{observe} that we have gained a factor of 2.1 in sensitivity by holding the source position fixed in our search. Note that a much greater improvement comes from knowing the binary candidate's period, as demonstrated in Figures \ref{fig:2d_hist} and \ref{fig:Mc_plot}.

With the framework developed in \citet{hasasia} we can construct detection sensitivity curves to estimate the PTA that will be required to detect or rule out the mass model presented in \citetalias{Iguchi2010}. The \texttt{hasasia} \citep{hasasia_code} package\footnote{\url{https://hasasia.readthedocs.io/}} allows us to construct these detection sensitivity curves using a straight forward matched filter statistic and to simulate PTA data with control over the number of pulsars, observing cadence, timing precision, and data length. Using this software to estimate an idealized signal-to-noise ratio (S/N) (see Eqn (79) in \citet{hasasia}), assuming the parameters in \citetalias{Iguchi2010} and using the pulsar noise parameters in \citet{11yr_data} we obtain $S/N=0.87$. 
We used this software to extend the baseline of the the existing 11-year NANOGrav data set by adding new data to the existing pulsars with a timing precision and cadence that matches recent data. We also augmented the PTA, adding new pulsars with timing precisions and cadences similar to those already in the array; we added pulsars for each projected year at a rate comparable to the current growth-rate of NANOGrav, which has been approximately 7 pulsars per year for the past three data sets.
%\footnote{Ongoing surveys for millisecond pulsars are regularly supplying good timing pulsars to the world's PTAs at a rate of several per year, as evidenced by the growth between the NANOGrav 5-, 9-, and 11-year data.}

We find that NANOGrav should be able to \rev{detect or} rule out the existence of a SMBHB in \src\ with the \citetalias{Iguchi2010} mass within five to eight years from the end of the data set considered here. However, while \texttt{hasasia} allows us to calculate the PTA's sensitivity to a CW at a specific sky location, it is unable to set other parameters (such as luminosity distance) as known due to electromagnetic information \rev{about the GW source} as is done in this work. As is discussed above, including source parameters that are electromagnetically derived to reduce the parameter space of the GW search allows for an increased sensitivity. Because of this, using electromagnetic information will likely allow us to accelerate this estimated timeline. To more reliably estimate this timescale, detailed simulation work will be necessary to quantify the improvement made by including electromagnetic information over typical searches.

Because the sensitivity of the array depends heavily on the observing baseline of each pulsar, the inclusion of additional data can help tremendously. Data of this sort are accessible through the IPTA \citep{dr2}, and followup analyses of \src\ by the international community could prove fruitful. 
This timeline to the PTA sensitivity required to confirm or deny \src\ as a SMBHB will be reduced with the more rapid addition of pulsars to the array, e.g., by adding more than 7 per year. This improvement will be accelerated if the newly included pulsars are near the sky location of \src, as, currently, there are few pulsars in the array that are very sensitive to \src. To accomplish this, pulsar searches should be undertaken near the sky locations of potential PTA targets to begin improving our sensitivity more rapidly. Some pulsars in this area of the sky can be included through use of data provided by the IPTA \citep{dr2}, showing once again that an international effort to detect \src\ could be worthwhile. 

In addition to the results for GWs from \src, our work has many implications for detection prospects of other binary candidates. As discussed in Section \ref{ftest} and shown in Figure \ref{fig:Mc_plot}, for \src, it was not until we widened our prior to span an order of magnitude in frequency space on either side of the target frequency that sensitivity was lost. For similar candidates, particularly those at similarly high orbital frequencies,
we presume that this result will hold. Therefore, as long as the sky location and luminosity distance of a potential target are known, a search of this type is worth attempting if at least an estimate of an orbital period can be obtained. We will caution that this improvement will differ depending on the sky location of the source, and that the amount of frequency-space that can be effectively searched with this method will be larger for higher-frequency sources. As can be seen in Figure \ref{fig:2d_hist}, it is the inclusion of samples at low frequencies that raise the upper limit. However, typical errors on binary periods are quite a bit smaller than the limit suggested here, meaning that this method should prove useful for most binary candidates. This method will also account for any frequency error due to unaccounted for frequency evolution of the SMBHB, which, in the case of a detection, would provide important constraints for evolutionary models. 

%% file: Conclusions.tex
In this work, we present a new method for performing multi-messenger searches for individual SMBHBs, using \src\ as a test case. \src\ was first identified as a binary candidate by \citetalias{Sudou2003}, and was first visited by PTAs in \citetalias{Jenet2004}, which ruled out the proposed binary model. In the intervening 15 years, a revised model was published by \citetalias{Iguchi2010} and PTA data and analysis methods have greatly improved. We used the NANOGrav 11-year data set, as well as the collaboration's flagship GW detection package, \eprise, to search for GWs from \src. Here, we are able to limit \src's chirp mass, at 95\% confidence, to $(1.65\pm0.02) \times 10^9~{M_\odot}$, a factor of 4.3 smaller than the limit set in \citetalias{Jenet2004}. However, we are unable to rule out the existence of a binary corresponding to the revised model proposed in \citetalias{Iguchi2010}. 

In addition to directly placing a limit on the chirp mass of \src\ for the published orbital period, we are able to quantify how much this multi-messenger approach increases our sensitivity over a typical `blind' PTA search. We have conducted a search on real data that includes GW frequency as a free parameter, and from this analysis, we learn that by including frequency constraints from electromagnetic binary source measurements to restrict the prior, we can gain approximately an order of magnitude in sensitivity when compared to a frequency-blind search spanning the whole PTA band. However, this drop in sensitivity does not occur until a relatively wide range of frequencies is searched over, meaning that this approach will be useful even for candidates with relatively poor constraints on their orbital periods.